\documentclass[prd,twocolumn,amsmath,showpacs,amssymb,floatfix,nofootinbib,10pt]{revtex4}
\usepackage{indentfirst}
\usepackage{tikz}
\usepackage{setspace}
\usepackage{amsfonts,color,xcolor,graphicx,amsfonts,multirow,amsmath}
\definecolor{bluee}{rgb}{0,0,1}
\usepackage[colorlinks=true, colorlinks=true, citecolor=bluee, linkcolor=bluee, urlcolor=bluee]{hyperref}
\usepackage{ulem}

\allowdisplaybreaks

\newcommand{\DS}[1]{/\!\!\!#1}
\definecolor{darkerblue}{rgb}{0.1, 0.1, 0.7}
\definecolor{darkergreen}{rgb}{0.1, 0.5, 0.1}
\definecolor{lightred}{rgb}{1, 0, 0}

\begin{document}
\title{Investigation for $D^+ \to \pi^+ \nu\bar\nu$ decay process within QCDSR approach}
\author{Yu Chen}
\address{Department of Physics, Guizhou Minzu University, Guiyang 550025, P.R.China}
\author{Hai-Bing Fu}
\email{fuhb@gzmu.edu.cn}
\author{Tao Zhong}
\email{zhongtao1219@sina.com}
\address{Department of Physics, Guizhou Minzu University, Guiyang 550025, P.R.China}
\author{Sheng-Bo Wu}
\address{Department of Physics, Guizhou Minzu University, Guiyang 550025, P.R.China}
\author{Dong Huang}
\address{Center of Experimental Training, Guiyang Institute of Information Science and Technology, Guiyang 550025, P.R.China}

\begin{abstract}
In the paper, we investigate the charmed meson rare decay process $D^+ \to \pi^+\nu\bar\nu$ by using QCD sum rules approach. Firstly, the pion twist-2 and twist-3 distribution amplitude $\xi$-moments $\langle\xi_{2;\pi}^n\rangle|_\mu$ up to 10th-order and $\langle \xi_{3;\pi}^{(p,\sigma),n}\rangle|_\mu$ up to fourth-order are calculated by using QCD sum rule under background field theory. After constructing the light-cone harmonic oscillator model for pion twist-2, 3 DAs, we get their behaviors by matching the calculated $\xi$-moments. Then, the $D\to \pi$ transition form factors are calculated by using QCD light-cone sum rules approach. The vector form factor at large recoil region is $f_+^{D\to\pi}(0) =  0.627^{+0.120} _{-0.080}$. By taking the rapidly $z(q^2,t)$ converging simplified series expansion, we present the TFFs and the corresponding angular coefficients in the whole squared momentum transfer physical region. Furthermore, we display the semileptonic decay process $\bar D^0 \to \pi^+ e\bar \nu_e$ differential decay widths and branching fraction with ${\cal B}(\bar D^0\to\pi^+e\bar\nu_e) = 0.308^{+0.155}_{-0.066} \times 10^{2}$. The $\bar D^0\to\pi^+e\bar\nu_e$ differential/total predictions for forward-backward asymmetry, $q^2$-differential flat terms and lepton polarization asymmetry are also given. After considering the non-standard neutrino interactions, the predictions for the $D^+ \to \pi^+ \nu\bar\nu$ branching fraction is ${\cal B}(D^+ \to \pi^+ {\nu }{\bar\nu}) = 1.85^{+0.93}_{-0.46}\times10^{-8}$.
\end{abstract}
\date{\today}

\maketitle

\newpage
\section{Introduction}
The remarkable success of Standard Model (SM) in describing all current experimental information suggests that the search for deviations from it should focus on either higher energy scales or small effects in low energy observables.\cite{Burdman:2001tf}. Normally, the flavour changing neutral current (FCNC) transitions in the SM are highly suppressed by the Glashow-Iliopoulos-Maiani (GIM) mechanism~\cite{Glashow:1970gm}. This GIM suppression has more strongly effective in charm sector compared to the down-type quarks in the bottom and strange sectors. Meanwhile, the suppression is responsible for the relatively small size of charm mixing and $C\!P$ violation in the charm system~\cite{Asner:2008nq,Saur:2020rgd,Wilkinson:2021tby}. So the FCNC processes of $D$-meson decays into charged lepton pairs are always totally overshadowed by long-distance contributions~\cite{Cappiello:2012vg, Aaij:2017iyr}. However, for $D$-meson FCNC decays into final states involving dineutrinos, such as $D^+\to \pi^+ \nu \bar \nu$, long-distance contributions become insignificant and the short-distance contributions from $Z$-penguin and box diagrams are dominant, which results in the branching fraction at the level of $10^{-16}$ in SM~\cite{Mahmood:2014dkp}. That makes $D$-meson FCNC decay involving dineutrinos a unique and clean probe to study the $C\!P$ violation in the charm sector~\cite{Bigi:2011em} and search for new physics beyond SM~\cite{Bause:2020xzj}.

On the experimental side, LHCb Collaboration reported an evidence for the breaking of lepton universality in bottom-quark FCNC decays to charged dielectrons and dimuons with a significance of 3.1$\sigma$~\cite{LHCb:2021trn}, which suggests the possible presence of new physics contributions in the lepton sector~\cite{European}. The neutral charmed meson FCNC decay into dineutrinos pair have been observed by the BESIII Collaboration, which provide the upper limits at $10\%$ confidence level for the $D^0 \to \pi^0 \nu \bar\nu$ branching fraction, {i.e. $2.1\times 10^{-4}$} in 2021~\cite{BESIII:2021slf}. This value is much larger than the SM prediction both from the long-distance and short distance. Non-standard neutrino interactions (NSIs)~\cite{Botella:1986wy, Valle:1987gv, Roulet:1991sm, Guzzo:1991hi, Bergmann:2000gp, Guzzo:2000kx, Guzzo:2001mi, Grossman:1995wx, Bergmann:1999rz, DeGouvea:2001mz,Davidson:2003ha}, which described by four fermion operators of the $(\bar\nu_\alpha\gamma\nu_\beta)(\bar f\gamma f)$ can narrow the gap between the experimental and the SM's predictions. It has been shown that the NSIs would be compatible with the oscillation effects along with some new features in various neutrino searches \cite{Johnson:1999ci, Gago:2001xg, Ota:2002na, Ota:2001pw, Campanelli:2002cc, Huber:2001de, Kitazawa:2006iq}. Meanwhile, NSIs are thought to be well matched with the oscillation effects, along with new features in neutrino searches~\cite{Barger:1991ae, Berezhiani:2001rs, Barranco:2005ps, Mangano:2006ar, Blennow:2007pu, Barranco:2007tz, Kopp:2007mi}.  The branching ratio of $D^+\to \pi^+ \nu\bar\nu$ could be at the level $10^{-8}$.  This approach can establish $D^+\to \pi^+ \nu \bar{\nu}$ direct connection with $\bar D^0 \to \pi^+ e \bar{\nu}_e$, which lead to the motivation in this paper.

The $D\to \pi $ transition form factors (TFFs) are the key component of semileptonic decays $\bar D^0\to \pi^+ e \bar\nu_e$. There are some researches dealing with the $D\to \pi $ TFFs both from experimental and theoretical side. Experimentally, the Belle~\cite{Belle:2006idb}, BESIII~\cite{BESIII:2015tql}, \textsc{Babar}~\cite{BaBar:2014xzf}, CLEO~\cite{CLEO:2009svp} Collaborations have measured the vector TFFs  value at large recoil region $f_+^{D\to\pi} (0)$. Meanwhile, the Lattice QCD also give this value~\cite{FermilabLattice:2004ncd}. Theoretically, the $D\to \pi $ TFFs can be calculated by using the QCD light-cone sum rule (LCSR)~\cite{Khodjamirian:2000ds}, light-front quark model (LFQM)~\cite{Verma:2011yw}, combined heavy meson and chiral Lagrangian theory HM$\chi$T~\cite{Fajfer:2004mv}. The LCSR approach is mainly extension on the light cone, which is expected to be valid at small and intermediate squared momentum transfer. After extrapolating the TFFs by using a suitable series expansion, one can get the vector and scalar TFFs in the whole physical region. Then one can make a comparison with other approaches. So in this paper, we mainly take the LCSR method to calculate the $D\to \pi$ TFFs.

Furthermore, the pion DAs with different twist structures are the key long-distance nonperturbative component in $D\to\pi$ TFFs LCSR expressions, which describes either contributions of the transverse motion of quarks (antiquarks) in the leading-twist components or contributions of higher Fock states with additional gluons and/or quark-antiquark pairs. Considering the contributions for each twist DA, the leading-twist and twist-3 DAs are dominant. The higher twists DAs contributions are highly suppressed by the energy scale and Borel parameters~\cite{Ball:2006wn,Ball:1998je}. For pion leading-twist DA, there have some predictions coming from theoretical group, such as the Lattice QCD (LQCD)~\cite{Bali:2019dqc}, DS model~\cite{Chang:2013pq} and QCD/AdS model~\cite{Ahmady:2018muv}. To the pion twist-3 DAs, the light-front quark model (LFQM)~\cite{Arifi:2023uqc} and QCDSR~\cite{Huang:2004tp,Huang:2005av,Braun:1989iv} have given the predictions. To get a more better ending-point behavior of pion twist-2, 3 DAs, we will calculated the $\xi$-moments and reconstruct the light-cone harmonic oscillator models.

The rest of paper are organized as follows. In Sec.~\ref{section:2}, we present the NSIs and branching fraction for $D^+\to \pi^+\nu_\ell \bar\nu_{\ell'}$, the $D\to \pi$ TFFs, pion twist-2 and twist-3 LCHO model and $\xi$-moment within QCDSR approach. In Sec.~\ref{section:3}, we give a detailed numerical and phenomenological analysis. Finally, a brief summary is provided in Sec.~\ref{section:4}

\section{Calculation Technology}\label{section:2}

An effective four fermion interactions including neutrinos, which is called NSIs, have the following Lagrangian
\begin{eqnarray}
{\cal L}^{\rm NSI}_{\rm eff} &=& - 2\sqrt{2} G_F \varepsilon^{fP}_{\ell \ell'}
(\bar\nu_\ell \gamma_\mu L \nu_{\ell'} )
(\bar f \gamma^{\mu} P f ) ,
\label{Eq:NSIL}
\end{eqnarray}
where $\ell$ and $\ell'$ stand for the light neutrino flavour. The symbol $f$ is the charged lepton or quark. $P=(L, R)$ with $L(R)=(1\mp \gamma_5)/2$ stand for the left or right operators. $\varepsilon^{fP}_{\ell \ell'}$ is the parameter for NSIs, which carry information about dynamics. The effective Hamiltonian governing the decays $D\to \pi \nu\bar\nu$, resulting from the $Z^0$-penguin and box-type contributions, can be written as~\cite{Chen:2007cn}
\begin{align}
{\cal H}_{\rm eff} &= \frac{G_F}{\sqrt 2} \frac{\alpha}{2\pi\sin^2\theta_W} \sum\limits_{\ell = e,\mu, \tau}[V_{cs}^*V_{cd} X_{\rm NL}^\ell + V_{ts}^*V_{td} X(x_t)]
\nonumber\\
&\times (\bar s d)_{V-A}(\bar\nu_\ell \nu_\ell)_{V-A}.
\end{align}
The $X_{\rm NL}^\ell$ is the charm quark contribution, and $X(x_t)$ is representing the loop integral of the top quark contribution. {\it i.e.} $X(x_t) = \eta_X x_t /8 \times [(x_t+2)/(x_t - 1) + (3x_t - 6)/(x_t - 1)^2 \ln x_t] $. The expressions $x_t = m_t^2/m_W^2$ and $(\bar f f')_{V-A} = \bar f \gamma_\mu (1- \gamma_5)f'$ have been adopted. The processes are dominated by short distance because long distance contributions are almost $10^{-3}$ less than short distance. The up-type quark in the loop will increase the branching ratios of these reactions. The NSIs in Eq.~\eqref{Eq:NSIL} can induce the transition $c\to u\nu_\ell \bar\nu_{\ell'}$ at one-loop level with a Feynman diagram, as we can be obtained~\cite{Chen:2007cn}
\begin{align}
H^{\rm NSI}_{c \to u\nu_\ell\bar\nu_{\ell'}}&= \frac{G_F}{\sqrt{2}}  \left( \frac{\alpha_{em}}{4\pi \sin^2\theta_W}
V_{cd} V^{*}_{ud} \varepsilon^{dL}_{\ell \ell'} \ln \frac{\Lambda}{m_W} \right)
\nonumber\\
&\times  (\bar\nu_\ell \nu_{\ell'})_{V-A} (\bar c u)_{V-A}.
\end{align}
Where $\theta _W$ is the Weinberg angle, $V_{cd}= 0.225$ and $V_{ud}^{*}=0.97370$ . Accordingly, the branching fraction for $D^+ \to \pi^+ \bar\nu_\ell\nu_{\ell'}$
is given, which leads to
\begin{align}
{\cal B}{(D^+ \to \pi^+ {\nu_\ell }{\bar\nu _{\ell '}})_{{\rm{NSI}}}} &= \bigg|V_{ud}^* \frac{{{\alpha _{\rm em}}}}{{4\pi {{\sin }^2}{\theta _W}}}\varepsilon _{\ell {\ell '}}^{dL}\ln \frac{\Lambda }{{{m_W}}}\bigg|^2
\nonumber\\
&\times {\cal B}(\bar D^0\to \pi^+ e \bar\nu_{e})
\end{align}
In order to study relevant physical observables, we adopt the explicit expression for the full differential decay width distribution of $ D\to \pi \ell \bar\nu_\ell$ as follows~\cite{Becirevic:2016hea,Cui:2022zwm}:
\begin{align}
\frac{d^2\Gamma(D\to \pi \ell \bar\nu_\ell)}{d\cos\theta dq^2} &=  a_{\theta_\ell}(q^2) + b_{\theta_\ell}(q^2) \cos\theta_\ell
\nonumber\\
& + c_{\theta_\ell}(q^2) \cos^2\theta_\ell
\label{Eq:dGamma}
\end{align}
 where the three $q^{2}$-dependent angular coefficient functions have the following expressions:
\begin{align}
a_{\theta_\ell}(q^2) &= {\cal N}_{\rm ew} \lambda^{3/2}\bigg(1 - \frac{m_\ell^2}{q^2}\bigg)^2 \bigg[|f^{D\to \pi}_+(q^2)|^2 + \frac{m_\ell^2}{q^2 \lambda}
\nonumber\\
&\times
\bigg( 1- { m_\pi^2 \over m_D^2 } \bigg)^2 |f^{D\to \pi}_0 (q^2)|^2\bigg],
\\
b_{\theta_\ell}(q^2)&= 2{\cal N}_{\rm ew} \lambda \bigg(1 - {m_\ell^2 \over q^2}\bigg)^2 {m_\ell^2 \over q^2}\bigg( 1-{m_\pi^2\over m_D^2 } \bigg)
\nonumber\\
&\times  {\rm Re} \left[ f^{D\to \pi}_{+}(q^2)  f^{D\to \pi\ast}_0(q^2) \right],
\\
c_{\theta_\ell}(q^2)&=-{\cal N}_{\rm ew} \lambda^{3/2} \bigg( 1 - {m_\ell^2 \over q^2} \bigg)^3  |f^{D\to \pi}_+(q^2) |^2.
\end{align}
From which, the electro-weak normalized coefficiency can be expressed as ${\cal N}_{\rm ew} = {G_F^2  |V_{cd}|^2   m_D^3 }/(192\pi^3)$ and $ \lambda(a, b, c) \equiv  a^2 + b^2 + c^2 - 2  (ab + ac +bc) $. In this paper, we take the shorthand notations for $\lambda \equiv \lambda (1,  m_\pi^2/m_D^2,  q^2/m_D^2 ) $ for convenience. In addition,  the helicity angle $\theta_\ell$ is defined as the angle between the $\ell^-$ direction of flight and the final-state meson  momentum in the dilepton rest frame. Thus in the massless lepton limit, we can observe two interesting algebra relations for the angular functions $b_{\theta_\ell}(q^2) =0$ and  $a_{\theta_\ell}(q^2) + c_{\theta_\ell}(q^2) =0$.

Furthermore, we should carry out a calculation for the $D\to \pi $ TFFs. In the first place, to derive LCSR expressions for the $D\to \pi$ TFFs, we will calculate take the standard correlator as the starting point, and then there's operator product expansion (OPE) near the light cone near zero. The vacuum-to-pion correlation function used to obtain the LCSR for the form factors of $D\to \pi $ transition is defined as:
\begin{align}
\Pi_\mu(p,q)=i\int d^4x e^{i q\cdot x}\langle \pi^+(p)|T\{j_\mu(x), j^\dag_5(0)\}|0\rangle,
\label{Eq:correlator}
\end{align}
with the two currents $j_\mu(x) = \bar u(x)\gamma_\mu c(x)$ and $j^\dag_5(0) = m_c\bar{c}(0)i\gamma_5 d(0)$. After taking the $c$-quark propagators into the correlation function and making operator product expansion, we can get the $D\to\pi$ TFFs OPE expression. On the other hand, after inserting hadronic states to the correlator Eq.~\eqref{Eq:correlator}, one can isolate the ground-state $D$-meson contributions in the dispersion relations for all three invariant amplitudes:
\begin{align}
\Pi^{\rm H}(p,q) &=\frac{2f^{D\to \pi}_+(q^2) m_D^2 f_{D}}{m_c(m_D^2 -(p+q)^2)}+\int_{s_0}^\infty ds\frac{\rho(s)}{s-(p+q)^2}
\nonumber\\
&+\text{subtractions}.
\\
\widetilde{\Pi}^{\rm H}(p,q) &=\frac{\tilde f^{D\to \pi}(q^2) m_D^2 f_D}{m_c(m_D^2 -(p+q)^2)}+\int_{s_0}^\infty ds\frac{\tilde\rho(s)}{s-(p+q)^2}
\nonumber\\
&+\text{subtractions}.
\label{Eq:Hadronic1}
\end{align}
The $D\to \pi $ form factors entering the residues of the $D$-meson pole in Eq.~\eqref{Eq:Hadronic1} are defined as:
$\langle\pi^+(p)|\bar{u} \gamma_\mu c |\bar D(p+q)\rangle = 2f^{D\to \pi}_+(q^2)p_\mu +\tilde f^{D\to \pi}q_\mu$
with $\tilde f^{D\to \pi}(q^2) = f^{D\to \pi}_+(q^2) + f^{D\to \pi}_{-}(q^2)$. Meanwhile the $f_D = \langle D |m_c\bar d i\gamma_5 d |0 \rangle/m_D^2$ is $D$-meson decay constant. With the help of quark-hadron duality, introducing the effective threshold parameter $s_0$. After the Borel transformation in the variable $(p+q)^2\to M^2$, the sum rules for  $D\to \pi $ form factors are obtained.
The LCSR for the vector form factor reads:
\begin{align}
f^{D\to \pi}_+(q^2)=\frac{e^{m_D^2/M^2}}{2m_D^2 f_D}&\Bigg[F_0(q^2,M^2,s_0)
\nonumber\\
&+\frac{\alpha_s C_F}{4\pi} F_1(q^2,M^2,s_0)\Bigg],
\\
\tilde f^{D\to \pi}(q^2) = \frac{e^{m_D^2/M^2}}{m_D^2 f_D}&\Bigg[\tilde F_0(q^2,M^2,s_0)
\nonumber\\
&+\frac{\alpha_s C_F}{4\pi} \tilde F_1(q^2,M^2,s_0)\Bigg],
\end{align}
where the LO expression for $F_0(q^2,M^2,s_0)$ and $\tilde F_0(q^2,M^2,s_0)$ have the following forms
\begin{widetext}
\begin{align}
&F_0(q^2,M^2,s_0)=m_c^2 f_\pi \int_{u_0}^1 du
e^{-\frac{m_c^2-\bar{u}q^2}{uM^2}}\bigg\{\frac{\phi_{2;\pi} (u,\mu)}u + \frac{\mu_\pi}{m_c}
\Bigg[\phi_{3;\pi}^p(u) + \bigg(\frac1{3u}-\frac{m_c^2+q^2}{6(m_c^2-q^2)} \frac d{du} \bigg)
\nonumber\\
&\qquad \qquad\times \phi_{3;\pi}^\sigma(u,\mu) \bigg]-2\bigg(\frac{f_{3;\pi}}{m_c f_\pi}\bigg) \frac{I_{3;\pi}(u)}{u} +\frac1{m_c^2-q^2}\bigg[-\frac{m_c^2 u}{4(m_c^2-q^2)} \frac{d^2 \phi_{4;\pi}(u)} {du^2} + u\psi_{4;\pi}(u)
\nonumber\\
&\qquad\qquad+ \int_0^u dv \psi_{4;\pi}(v)-I_{4;\pi}(u)\bigg]\bigg\}
\label{Eq:F0}
\\
&\tilde F_0(q^2,M^2,s_0)=m_c^2 f_\pi \int_{u_0}^1 du
e^{-\frac{m_c^2-\bar{u}q^2}{uM^2}}\bigg\{\frac{\mu_\pi}{m_c}\bigg(\frac{\phi_{3;\pi}^p(u)}{u}+ \frac1{6u} \frac{d\phi_{3;\pi}^\sigma(u)}{du}\bigg) + \frac1{m_b^2 - q^2} \psi_{4;\pi}(u)\bigg\}.
\label{Eq:tildeF0}
\end{align}
\end{widetext}
where $\mu_\pi={m_\pi^2}/{(m_u+m_d)}$, $u_0=(m_c^2-q^2)/(s_0-q^2)$~\cite{A.Khodjamirian09}. $s_0$ is the effective threshold parameter, $f_D$ and $f_\pi$ is decay constants of $D$ and $\pi$-mesons, $m_c$ is the charm quark mass. $\phi_{2;\pi}(u,\mu)$ and $\phi_{3;\pi}^{p}(u,\mu)$, $\phi_{3;\pi}^{\sigma}(u,\mu)$ are the pionic twist-2 and twist-3 DA's, respectively. While, the allowable physical range is $0 \leq q^2\leq  (m_D - m_\pi)^2 \approx 2.9{\rm GeV^2}$. The NLO correction $F_1(q^2,M^2,s_0)$ is mainly comes from Refs.~\cite{Duplancic:2008ix,Hu:2021zmy}.

Furthermore, in dealing with the pion different twist DAs especially twist-2 and twist-3 DAs, at processes related typical scale, we can take the Brodsky-Huang-Lepage (BHL) description \cite{Huang:1994dy}, and it is the light-cone harmonic oscillator model (LCHO model) of the pion twist-2 and twist-3 WF~\cite{Wu:2010zc, Wu:2011gf}. The wavefunction for the twist-2 LCDA can be expressed as $\Psi_{2;\pi}(x,\textbf{k}_\bot) = \sum_{\lambda_1\lambda_2} \chi_{2;\pi}^{\lambda_1\lambda_2}(x,\textbf{k}_\bot) \Psi^R_{2;\pi}(x,\textbf{k}_\bot)$
with $\textbf{k}_\bot$ is the pionic transverse momentum, $\lambda_1$ and $\lambda_2$ are the helicities of the two constituent quarks. $\chi_{2;\pi}^{\lambda_1\lambda_2}(x,\textbf{k}_\bot)$ replace the spin-space wavefunction (WF) that based on the Wigner-Melosh rotation, whose preformance for different $\lambda_1\lambda_2$ are exhibited, which can also been seen in Refs.~\cite{Cao:1997hw,Huang:2004su,Wu:2005kq}.
$\Psi^R_{2;\pi}(x,\textbf{k}_\bot) = A_{2;\pi} \varphi_{2;\pi}(x) \exp [ -(\textbf{k}^2_\bot + m_q^2)/(8\beta_{2;\pi}^2 x\bar x)]$ with $\bar{x} = (1-x)$. $A_{2;\pi}$ is the normalization constant, $\textbf{k}_\bot$-dependence part of the spatial WF $\Psi^R_{2;\pi}(x,\textbf{k}_\bot)$ comes from  quark model of pion and confirm the WF's transverse distribution and get through the harmonious parameter $\beta_{2;\pi}$. Here we take $m_q =250~{\rm MeV}$ in this paper. Combine with pionic leading-twist DA and WF through the relationship:
\begin{align}
\phi_{2;\pi}(x,\mu) =\frac{2\sqrt{6}}{16\pi^3f_\pi} \int_{| \mathbf{k}_\bot |^2 \leq \mu^2} d^2\mathbf{k}_\bot \Psi_{2;\pi}(x,\mathbf{k}_\bot),\label{Eq:DAandWF}
\end{align}
After taking the spin-space wavefunction and spatial wavefunction into the above formula, one can get the full LCHO expression for the pion leading-twist DA:
\begin{align}
&\phi_{2;\pi}(x,\mu) = \frac{\sqrt{3} A_{2;\pi} m_q \beta_{2;\pi}}{2\pi^{3/2}f_\pi} \sqrt{x\bar x} \varphi_{2;\pi}(x)
\nonumber\\
&\qquad\times  \left\{ \textrm{Erf}\left[ \sqrt{\frac{m_q^2 + \mu^2}{8\beta_{2;\pi}^2 x\bar x}} \right]-  \textrm{Erf}\left[ \sqrt{\frac{m_q^2}{8\beta_{2;\pi}^2 x\bar x}} \right] \right\}.
\label{DA_model}
\end{align}
with $\varphi_{2;\pi}(x) = [x\bar x]^{\alpha_{2;\pi}} [ 1 + \hat{a}^{2;\pi}_2 C_2^{3/2}(\xi) ]$ with $\xi = (2x-1)$. ${\rm Erf}(x) = 2\int^x_0 e^{-t^2} dx/{\sqrt{\pi}}$ is the error function~\cite{Zhong:2021epq}.  Meanwhile, the $\hat{a}^{2;\pi}_2$ are the second-order Gegenbauer moment, and the parameters $\alpha_{2;\pi}$ and $\hat{a}^{2;\pi}_2$  can be determined by fitting the moments $\langle\xi^n_{2;\pi}\rangle|_\mu$ directly through the method of least squares.

Meanwhile, the two pion twist-3 LCDAs can also be related to its wavefunction by using the formula
\begin{align}
\phi_{3;\pi}^{p,\sigma}(x,\mu) = \frac{1}{16\pi^3} \int_{| \mathbf{k}_\bot |^2 \leq \mu^2} d^2\mathbf{k}_\bot \Psi_{3;\pi}^{p,\sigma}(x,\mathbf{k}_\bot),
\end{align}
Here, we take the LCHO model for pion twist-3 wavefunction. Its basic idea is the longitudinal behavior is dominated by the first two Gegenbauer moments and transverse momentum dependence the BHL prescription. In this approach, the twist-3 DA will have a better end-point behavior other than asymptotic one shall. Then, we found that the transverse momentum dependence is just in the exponential form of the off-shell energy of the constituent quarks, which agrees with Brodsky and Teramond's holographic model that is obtained by using the Anti-de Sitter (AdS) conformal field theory correspondence \cite{wfbrodsky1, Brodsky:2007hb, Brodsky:2008pg}. The detail discussion can be seen in our previous work~\cite{Zhong:2011jf}. Then we can obtain:
\begin{align}
\phi_{3;\pi}^{p,\sigma}(x,\mu)&=\frac{A_{3;\pi}^{p,\sigma} (\beta_{3;\pi}^{p,\sigma})^2}{2\pi^2} \varphi_{3;\pi}^{p,\sigma}(x)\textrm{exp} \left[ -\frac{m_q^2}{8  (\beta_{3;\pi}^{p,\sigma})^2 x\bar x} \right]
\nonumber\\
&\times  \left\{ 1 - \textrm{exp} \left[ -\frac{\mu^2}{8  (\beta_{3;\pi}^{p,\sigma})^2 x\bar x}\right] \right\},  \label{Eq:phi3psigma}
\end{align}
with $\varphi_{3;\pi}^p(x) = 1+B_{3;\pi}^p C^{1/2}_2(\xi) + C_{3;\pi}^p C^{1/2}_4(\xi)$ and $\varphi_{3;\pi}^\sigma(x) = 1+B_{3;\pi}^\sigma C^{3/2}_2(\xi) + C_{3;\pi}^\sigma C^{3/2}_4(\xi)$. Next, we will face to the four free parameters in the LCHO model of pion twist-2 and twist-3 DAs. The first constraint is the twist-2, 3 DAs are normalized to 1. The second one is the average value of the squared transverse momentum is taken to be $(\langle{\bf k}_\bot^2\rangle_{2;\pi})^{1/2} = (\langle{\bf k}_\bot^2\rangle_{3;\pi}^{p,\sigma})^{1/2} = 0.35~{\rm GeV}$~\cite{Guo:1991eb}. Then, to determine the remaining two parameters, we need to calculate the pion DA $\xi$-moments which have the following definitions
\begin{align}
&\langle\xi^n_{2;\pi}\rangle|_\mu = \int_0^1 dx (2x-1)^n \phi_{2;\pi} (x,\mu),
\\
&\langle \xi_{3;\pi}^{(p,\sigma),n}\rangle|_\mu =  \int_0^1 dx (2x-1)^n \phi_{3;\pi}^{(p,\sigma)}(x,\mu).
\end{align}
So our next task is to calculate pion twist-2, 3 DA moments by using QCD sum rule method.

In order to derive the $\xi$-moments of $\pi$-meson twist-2 and twist-3 DAs $\phi_{2;\pi}(x)$ and $\phi_{3;\pi}^{p,\sigma}(x)$, we take the following correlation functions
\begin{align}
\Pi_{2;\pi} (z,q) &=  i \int d^4 x e^{iq \cdot
x} \langle 0|{\rm  T} \{J_{2;\pi}^n(x), J_{2;\pi}^{0\dag} (0) \}| 0\rangle, \nonumber\\
\Pi_{3;\pi}^{p,\sigma} (z,q) &= -i \int d^{4} x e^{iq \cdot
x} \langle 0|{\rm  T} \{J_{3;\pi}^{(p,\sigma),n}(x), J_5^{0\dag}(0) \}| 0\rangle,
\end{align}
where the twist-2 DA currents are $J_{2;\pi}^n (x) = \bar{d}(x) \DS{z}\gamma_{5} ( i z \cdot
\tensor{D})^{n} u(x)$ and $J_{2;\pi}^0 (x) = \bar{d}(x) \DS{z}\gamma_{5}  u(x)$. The twist-3 DA currents are $J_{3;\pi}^{p,n} (x) = \bar{d}( x ) \gamma_{5} ( i z \cdot
\tensor{D})^{n} u(x)$, $J_{3;\pi}^{\sigma,n}(x) = \bar{d}(x) \sigma_{\mu\nu} \gamma_5 ( i z \cdot \tensor{D} )^{n+1} u(x)$ and $J_5^{0\dag}(0) = \bar{u}( 0 ) \gamma_{5} d(0)$. The correction functions can also be translated into
\begin{align}
&\Pi_{2;\pi} (p,q) = (z \cdot q)^{n+2} I^{(n,0)}_{2;\pi} (q^2),
\nonumber\\
&\Pi_{3;\pi}^{p} (p,q) = (z \cdot q)^{n} I^{p,(n,0)}_{3;\pi} (q^2),
\nonumber\\
&\Pi_{3;\pi}^\sigma (p,q) = - i (q_\mu z_\nu - q_\nu z_\mu) (z \cdot q)^n I^{\sigma, (n,0)}_{3;\pi} (q^2).
\label{Eq:twist-3correlator}
\end{align}
After adopt the traditional QCD sum rule approach, we can get the following  analytical expressions for the $\xi$-moments for twist-2 and twist-3 pion DAs.
\begin{widetext}
\begin{eqnarray}
\frac{\langle\xi^n_{2;\pi}\rangle|_\mu \langle \xi^0_{2;\pi}\rangle|_\mu f_{\pi}^2}{M^2 e^{m_\pi^2/M^2}} &=& \frac{3}{4\pi^2(n+1)(n+3)}  \Big( 1 - e^{-s_{2;\pi}/M^2} \Big) + \frac{(m_d + m_u) \langle\bar qq \rangle }{(M^2)^2}
+ \frac{\langle \alpha_sG^2\rangle }{(M^2)^2} \frac{1 + n\theta(n-2)}{12\pi(n+1)}
\nonumber\\[1ex]
&-& \frac{(m_d + m_u)\langle  g_s\bar q\sigma TGq \rangle }{(M^2)^3}\frac{8n+1}{18} + \frac{\langle g_s\bar qq\rangle ^2}{(M^2)^3} \frac{4(2n+1)}{81} - \frac{\langle g_s^3fG^3\rangle }{(M^2)^3}\frac{n \theta(n-2)}{48\pi^2}
\nonumber\\[1ex]
&+& \frac{\langle g_s^2\bar qq\rangle ^2}{(M^2)^3} \frac{2+\kappa^2}{486\pi^2} \Big\{-2(51n+ 25)\Big(-\ln \frac{M^2}{\mu^2} \Big) + 3(17n+35) + \theta(n-2)\Big[ 2n
\nonumber\\
&\times&\Big(\!-\ln \frac{M^2}{\mu^2} \Big) + \frac{49n^2 +100n+56}n - 25(2n+1)\Big[ \psi\Big(\frac{n+1}{2}\Big) - \psi\Big(\frac{n}{2}\Big) + \ln4 \Big] \Big]\Big\}.
\label{Eq:twist2xinxi0}
\\
\frac{ \langle \xi_{3;\pi}^{p,n}\rangle|_\mu \langle \xi_{3;\pi}^{p,0} \rangle|_\mu f^2_\pi \mu_\pi^2}{M^4 e^{m^2_\pi / M^2}} &=&\frac{3}{8 \pi^2} \frac{1}{2n+1}\bigg [ 1 - \bigg( 1 + \frac{s^p_{3;\pi}}{M^2} \bigg) e^{-s_{3;\pi}^{p} / M^{2}} \bigg] + \frac{2n-1}{2}\frac{(m_u + m_d) \langle \bar q q \rangle }{M^4}
\nonumber\\
& +& \frac{2n+3}{24\pi} \frac{ \langle \alpha_s G^2 \rangle}{M^4} + \frac{16\pi}{81} [21+8n(n+1)] \frac{\langle \sqrt{\alpha_s} \bar q q \rangle^2}{M^6}
\label{Eq:twist3pxinxi0}
\\
\frac{\langle \xi_{3;\pi}^{\sigma,n}\rangle|_\mu  \langle \xi_{3;\pi}^{p,0} \rangle|_\mu  f^2_\pi \mu_\pi^2}{M^4 e^{m^2_{\pi} / M^2}} &=&  \frac{3}{2n+1}\bigg\{ \frac{3}{8 \pi^2} \frac{1}{2n+3}\bigg[ 1 -\bigg ( 1 + \frac{s^{\sigma}_{3;\pi}}{M^2}\bigg) e^{-s^{\sigma}_{3;\pi} / M^{2}} \bigg]+ \frac{2n+1}{2}\frac{(m_u + m_d) \langle \bar q q \rangle}{M^4} \nonumber\\
& +& \frac{2n+1}{24\pi} \frac{ \langle\alpha_s G^2 \rangle}{M^4} + \frac{16\pi}{81} (8n^2 - 2) \frac{\langle \sqrt{\alpha_s} \bar q q \rangle^2}{M^6} \bigg\},
\label{Eq:twist3sigmaxinxi0}
\end{eqnarray}
\end{widetext}
When taking order $0$th order $\xi$-moment normalized to be 1 directly, which is been adopted by many QCD sum rule,  there will have extra deviation to predicted $\langle\xi^n_{2;\pi}\rangle|_\mu$ and $\langle \xi_{3;\pi}^{(p,\sigma),n}\rangle|_\mu$. Thus we keep the $\langle\xi^0_{2;\pi}\rangle|_\mu$ and $\langle \xi_{3;\pi}^{(p,\sigma),0}\rangle|_\mu$ in Eqs.~\eqref{Eq:twist2xinxi0}-\eqref{Eq:twist3sigmaxinxi0}. The detailed  derivation for Eq.~\eqref{Eq:twist2xinxi0} and Eqs.~(\ref{Eq:twist3pxinxi0}, \ref{Eq:twist3sigmaxinxi0}) can be seen in Ref.~\cite{Zhong:2021epq} and Ref.~\cite{Zhong:2011jf}, respectively.

\section{numerical analysis}\label {section:3}

In order to perform the phenomenological analysis, the following input parameters need to be taken. The pole mass of $c$-quark is $m_c=1.50\pm0.05~{\rm GeV}$~\cite{Fu:2013wqa}.  The initial and final meson masses are $m_D=1869.66~{\rm MeV}$ and $m_\pi = 139.57039 \pm 0.00017~{\rm MeV}$, while the pion decay constant $f_\pi = 130.2\pm1.2~{\rm MeV}$ are taken from Particle Data Group (PDG)~\cite{PDGnew}. The $D$-meson decay constants are taken as $f_D = 0.163^{+0.017}_{-0.021}~{\rm GeV} $. The values of non-perturbative vacuum condensates up to six-dimension are taken as follows~\cite{Shifman:1980ui, Colangelo:2000dp, Narison:2014ska},
\begin{align}
&\langle \alpha_s G^2 \rangle = 0.038\pm0.011~{\rm GeV}^4,   \nonumber\\
&\langle g_s^3fG^3\rangle  = 0.045\pm0.007~{\rm GeV}^6,   \nonumber\\
&\langle g_s\bar qq\rangle ^2 = (2.082_{-0.697}^{+0.734})\times 10^{-3} ~{\rm GeV}^6,   \nonumber\\
&\langle g_s^2\bar qq\rangle ^2 = (7.420_{-2.483}^{+2.614})\times 10^{-3}~{\rm GeV}^6, \nonumber\\
&\langle q\bar q\rangle  = (-2.417_{-0.114}^{+0.227})\times 10^{-2}~{\rm GeV}^3, \nonumber\\
&\sum {\langle g_s^2\bar qq\rangle ^2}   = (1.891_{ - 0.633}^{ + 0.665})\times{10^{ - 2}}~{\rm GeV}^6.
\end{align}
The quark-gluon mixture condensate $\langle g_s\bar q\sigma TGq\rangle= m_0^2\langle \bar qq\rangle$ with $m_0^2 = 0.80 \pm 0.02{\rm GeV}^2$. Due to the $D\to \pi$ processes typical scale is $\mu_k = (m_D^2-m_c^2)\sim1.1~{\rm GeV}$. The renormalization group equations (RGEs) should be used to running the quark masses and each vacuum condensates appearing in the BFTSR from
the initial scale $\mu_0 = 1~{\rm GeV}$ to the typical scale $\mu_k$. In QCD sum rule approach, the continuum threshold $s_0$ and Borel parameter $M^2$ are the two important parameters which should be determined strictly. In this paper, we take the following four criteria:
\begin{itemize}
\item The continuum contributions are less than $45\%$ of the total results;
\item The contributions from the dimension-six condensates do not exceed $5\%$;
\item We require the variations of $\langle\xi_{2;\pi}^n\rangle|_\mu$ within the Borel window to be less than $10\%$
\end{itemize}
\begin{figure}[t]
\includegraphics[width=0.4\textwidth]{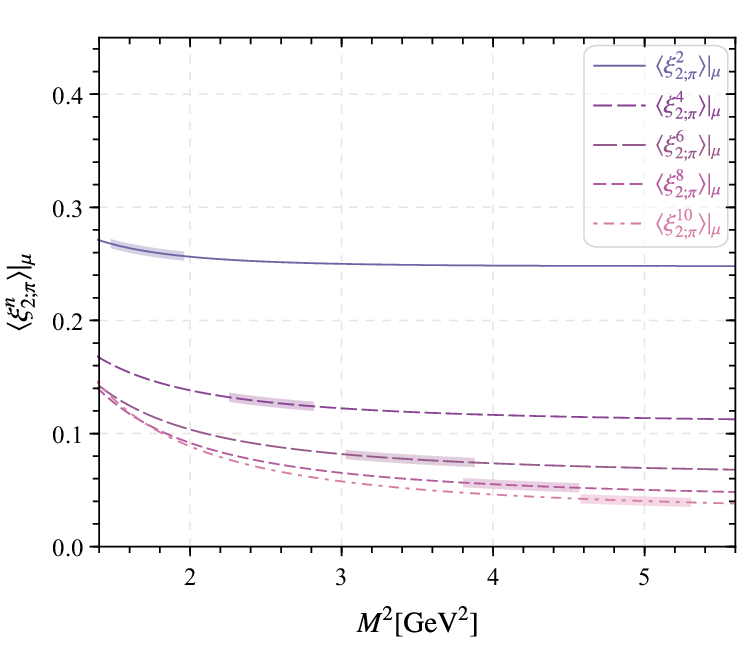}
\caption{Moments $\langle\xi^{n}_{2;\pi}\rangle|_\mu$ up to $n=(2,4,\cdots,10)$-order level versus the Borel parameter $M^2$. The shaded bands stand for the corresponding Borel windows.}\label{fig:xi}
\end{figure}
Based on the above criteria, we determine the value of continuum threshold parameter $s_{0}^\pi$ for pion leading-twist DA by using the 0th-order $\xi$-moments normalization, {\it i.e.} $\langle\xi^{0}_{2;\pi}\rangle|_\mu = 1$. So we have $s_0^\pi=1.05 {\rm GeV^2}$ for $n=(2,4,6,8,10)$-order moments. Furthermore, the allowable region for Borel parameters (also called Borel window) for each order $\xi$-moments are present in Fig.~\ref{fig:xi}, where the shaded region stand for the Borel windows. From the figure we can see that the Borel window increases with the increase of $n$th-order, which are all larger than 1. The values of $\langle\xi_{2;\pi}^n\rangle|_\mu$ are decreases with the increase of order $n$, which can also be seen in our previous work~\cite{Zhong:2021epq}.
\begin{figure}[t]
 \includegraphics[width=0.4\textwidth]{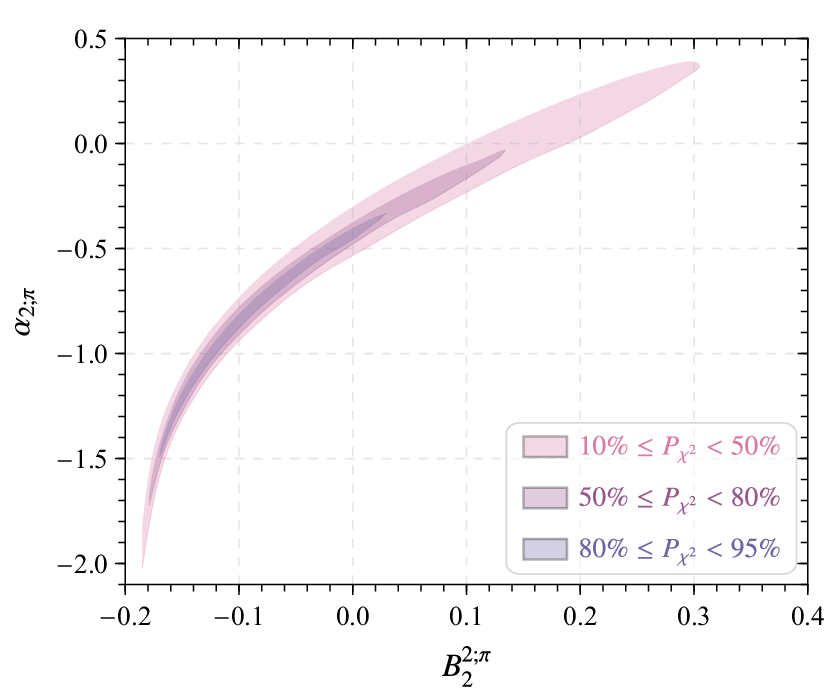}
\caption{The relationship curve between parameters $\alpha _{2;{\pi}}$, $B^{2;{\pi}}_{2}$ and goodness of fit $P_{\chi_{\min}^2}$ at factorization scale $\mu_k$.}    \label{fig:fitting}
\end{figure}
\begin{table}[t]
\footnotesize
\begin{center}
\caption{The ratios of the continuum states' and the dimension-six condensates' contributions over the total moments of $\pi$-meson twist-2 LCDA $\langle\xi_{2;\pi}^n\rangle|_\mu$ with $n=(2,4,6,8,10)$ within the determined Borel windows. The abbreviations ``Con.'' and ``Six.'' stand for the continuum and dimension-six contributions, respectively.}
\label {tab:m2}
\begin{tabular}{l c c c c c c}
\hline
$n$~~~~~~~~~~~~~~ & Con. & ~~~~~~~~~~~~~~~Six.~~~~~~~~~~~~~~~   & $\langle\xi_{2;\pi}^n\rangle|_\mu$                \\  \hline
2    & $ < 35\%$  & $ < 5\%$ &  $0.267\pm0.012$\\
4    & $ < 35\%$  & $ < 5\%$ &  $0.135\pm0.009$\\
6    & $ < 40\%$  & $ < 5\%$ &  $0.085\pm0.008$\\
8    & $ < 40\%$  & $ < 5\%$ &  $0.062\pm0.006$\\
10   & $ < 45\%$  & $ < 5\%$ &  $0.048\pm0.005$   \\  \hline
\end{tabular}
\end{center}
\end{table}
Then, the first five pion leading-twist DA $\xi$-moments {\it i.e.} $\langle\xi_{2;\pi}^n\rangle|_\mu$ with $n=(2,4,6,8,10)$ combing with the continuum states contributions and dimension-six contributions are listed in Table~\ref{tab:m2}. When $n=(2,4,6,8,10)$, the continuum contributions are set to be less than $35\%$, $35\%$, $40\%$, $40\%$, $45\%$, respectively. The dimension-six condensates' contributions to be less than $5\%$ for all the order of $\langle\xi_{2;\pi}^n\rangle|_\mu$.

\begin{figure}[t]
\begin{center}
\includegraphics[width=0.4\textwidth]{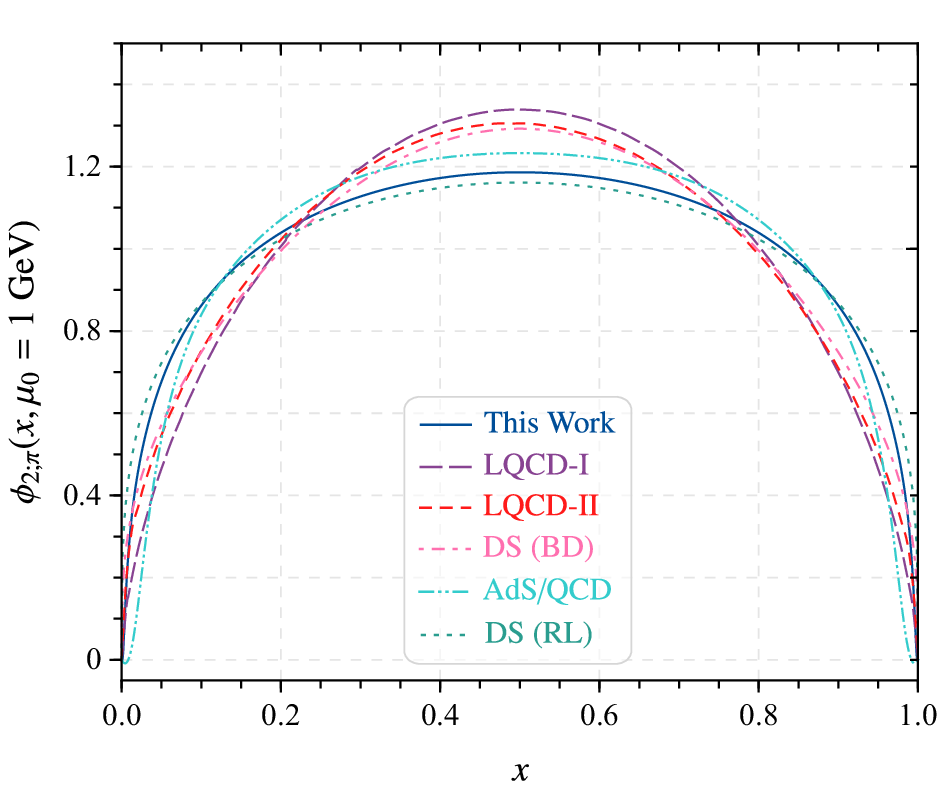}
\end{center}
\caption{The pion leading-twist DA curves of our prediction. As a comparison, we also present the LQCD~\cite{Bali:2019dqc}, DS model with BD and RL scheme~\cite{Chang:2013pq} and QCD/AdS model~\cite{Ahmady:2018muv} as a comparison.} \label{Fig:DAtwist2}
\end{figure}

In Fig.~\ref{fig:fitting}, the relationship between fitting parameters $\alpha _{2;{\pi}}$, $B^{2;{\pi}}_{2}$ and the goodness of fit at scale $\mu_k$ is plotted. The shaded bands stand for the corresponding Borel windows. The depth of color in the shaded band represents the degree of goodness of fit. The deeper the color, the higher the goodness of fit. When the range of goodness of fit is $80\%  \le P_{\chi^2_{\min}} \le 96\% $, the effect for goodness of fit is the best.
\begin{figure}[t]
\begin{center}
\includegraphics[width=0.4\textwidth]{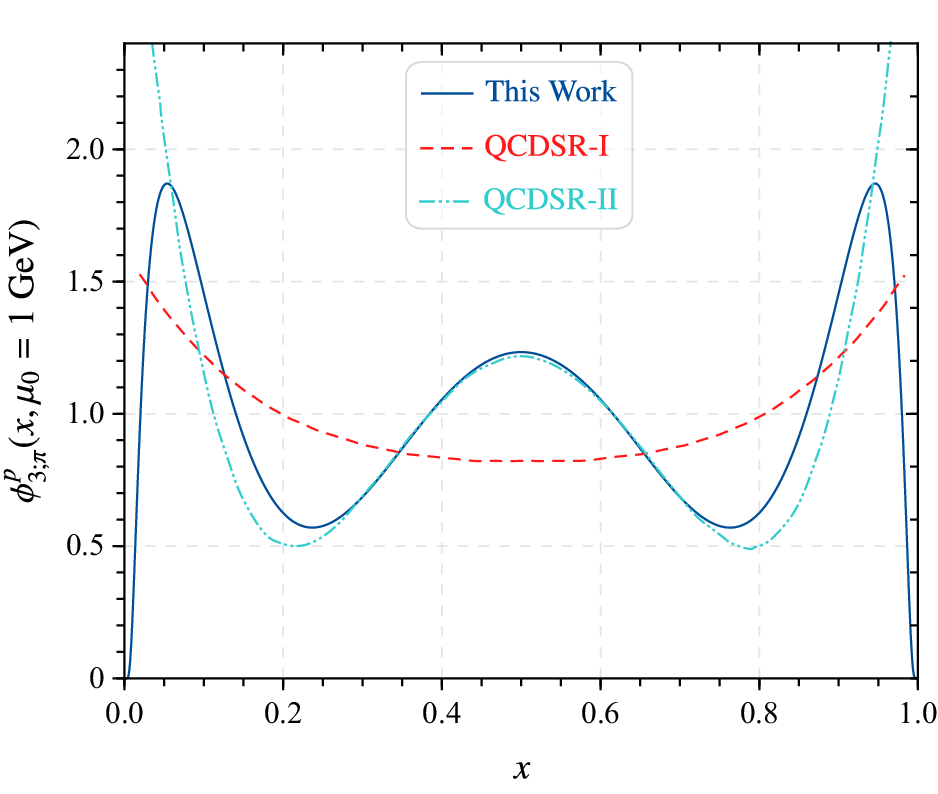}
\includegraphics[width=0.4\textwidth]{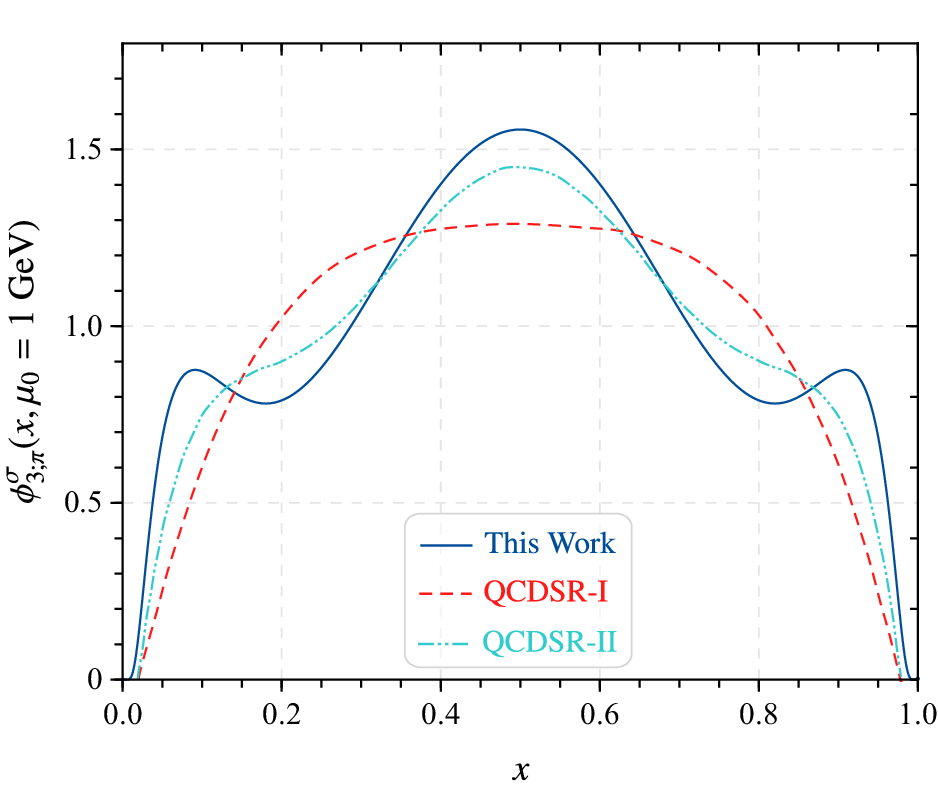}
\end{center}
\caption{The pion twist-3 DAs $\phi_{3;\pi}^{p}(x,\mu_0)$ and $\phi_{3;\pi}^{\sigma}(x,\mu_0)$ in this work. Meanwhile, predictions coming from QCDSR-I~\cite{Ball:1998je}, and QCDSR-II~\cite{Braun:1989iv} are also given as a comparison.} \label{Fig:DAstwist3}
\end{figure}
The obtained optimal values of the model parameters $\alpha_{2;\pi}$, $\hat{a}^{2;\pi}_2$ and $\beta_{2;\pi}$ at scale $\mu_0 =1~{\rm GeV}$ are
\begin{align}
&A_{2;\pi} = 3.659~{\rm GeV}^{-1}, && \alpha_{2;\pi} = -1.010,
\nonumber\\
& \hat{a}^{2;\pi}_2 = -0.126, && \beta_{2;\pi} = 0.727~{\rm GeV}.
\end{align}
The corresponding behaviors of pion leading-twist DAs $\phi_{2;\pi}(x,\mu_0)$ at initial scale $\mu_0 = 1~{\rm GeV}$ of our LCHO predictions is shown in Fig.~\ref{Fig:DAtwist2}. As a comparison, the model in literature for $\pi$-meson leading-twist DA $\alpha_{2;\pi}$ such as LQCD model~\cite{Bali:2019dqc}, DS model~\cite{Chang:2013pq}, QCD/AdS model~\cite{Ahmady:2018muv}. From the Fig.~\ref{Fig:DAtwist2}, one can find that our present prediction for $\phi_{2;\pi}$ is closely to DS model and QCD/AdS model.

As for the pion twist-3 DA $\xi$-moments, we take the continuum threshold $s_{3;\pi}^p = s_{3;\pi}^\sigma = 1.69(10)~{\rm GeV}^2$ and $\langle \xi_{3;\pi}^{p,0} \rangle|_\mu = 1$. After followed by the traditional criteria in determine the Borel window, we can get the first two order moments, which can be found in our previous work~\cite{Zhong:2011jf}. With the resultant $\langle \xi_{3;\pi}^{(p,\sigma),(2,4)}\rangle|_{\mu}$, we calculated the wavefunction parameters at initial scale $\mu_0 = 1~{\rm GeV}$ have the following values
\begin{align}
&A_{3;\pi}^{p} = 78.72~{\rm GeV}^{-2}, && B_{3;\pi}^{p} = 0.476,
\nonumber\\
& C_{3;\pi}^{p} = 0.943, && \beta_{3;\pi}^{p} = 0.608~{\rm GeV},\\
\nonumber\\
&A_{3;\pi}^{\sigma} = 160.5 {\rm GeV}^{-2}, && B_{3;\pi}^{\sigma} = -0.027,
\nonumber\\
& C_{3;\pi}^{\sigma} = 0.933, && \beta_{3;\pi}^{\sigma} = 0.424~{\rm GeV}.
\label{Eq:Parameters_twist-3DA}
\end{align}
The corresponding behaviors of pion twist-3 DAs $\phi_{2\pi}$ of our LCHO predictions is shown in Fig.~\ref{Fig:DAstwist3}. As a comparison, we also present other QCDSR results from literature for $\pi$-meson twist-3 DAs are also present, which is labeled as  QCDSR-I~\cite{Ball:1998je} and QCDSR-II~\cite{Braun:1989iv}.  Different with twist-2 DA, the pion twist-3 DAs especially for the $\phi_{3;\pi}^p(x)$ have larer discripency with other theoretical predictions. As can be seen in the upper panel of Fig.~\ref{Fig:DAstwist3}, our predictions have agreement with the QCDSR-II in the meddle $x$-region, {\it i.e.} $0.3\leq x\leq 0.7$, but have large difference in the region $0 \leq x < 0.3$ and $0.7 < x \leq 1$. For the $\phi_{3;\pi}^\sigma(x)$, the curve of our prediction have the same tendency with QCDSR-I. As for the twist-4 DAs, we take the expressions and input parameters from Ref.~\cite{Duplancic:2008ix}.

\begin{table}[b]
\caption {Comparison of theoretical predictions for the form factors $f_+^{D\to\pi}(0)$.}
\label{Tab:TFFfp0}
\begin{tabular}{ll}
\hline
References~~~~~~~~~~~~~~~~~~~~~~~~~~~&$ f_+^{D\to\pi}(0)$\\\hline
This work&$0.627^{+0.120} _{-0.080}$\\
Belle~\cite{Belle:2006idb}& $0.624\pm 0.020\pm0.030$\\
BESIII~\cite{BESIII:2015tql}&$0.637\pm 0.008\pm0.004$ \\
BES~\cite{BES:2004rav}&$0.730\pm 0.14\pm0.0060$ \\
BaBar~\cite{BaBar:2014xzf}&$0.610\pm0.020\pm0.005$\\
CLEO~\cite{CLEO:2009svp}&$0.640\pm 0.03\pm 0.06$\\
LQCD~\cite{FermilabLattice:2004ncd}&$0.640\pm0.03\pm0.06$\\
LCSR~\cite{Ball:2006yd}&$0.630\pm0.110$\\
\hline
\end{tabular}
\end{table}
\begin{figure}[t]
\begin{center}
\includegraphics[width=0.4\textwidth]{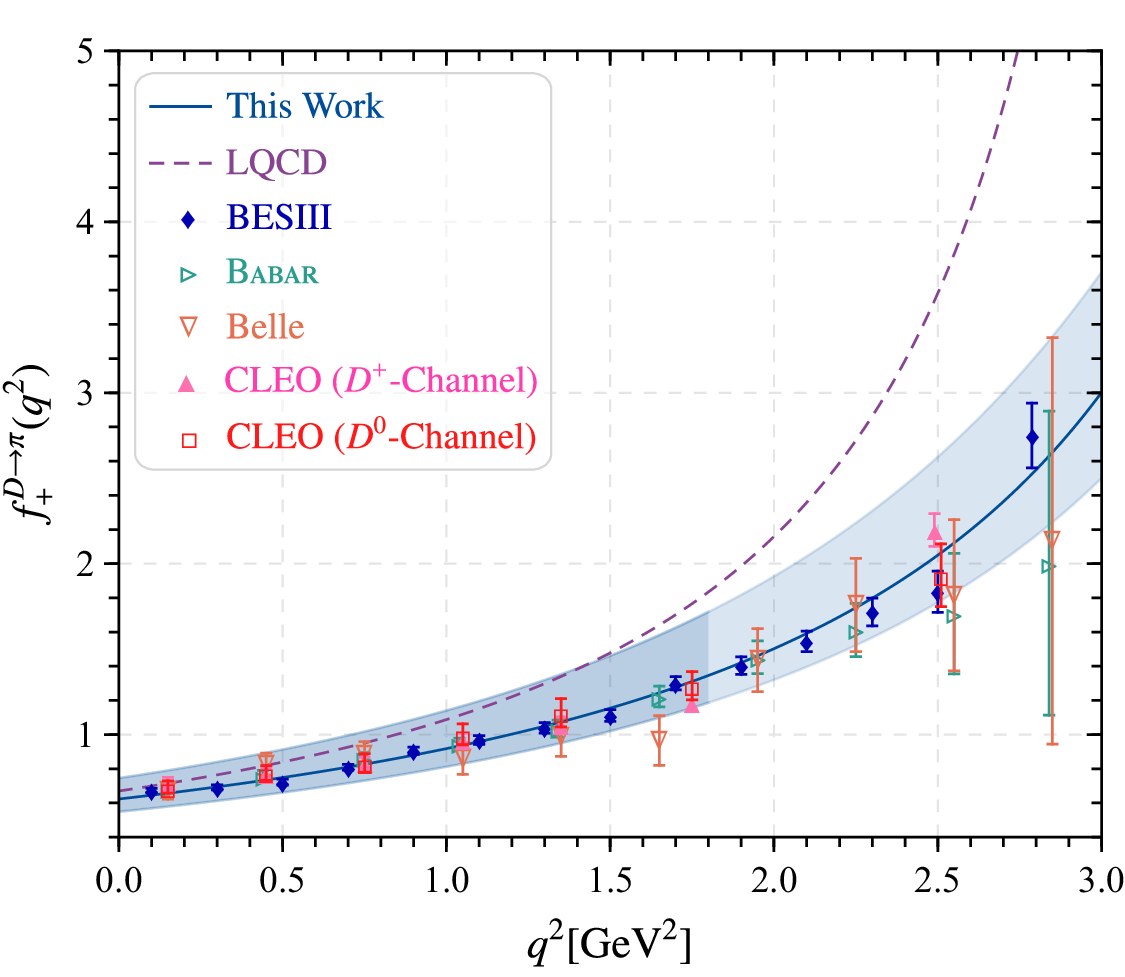}
\includegraphics[width=0.4\textwidth]{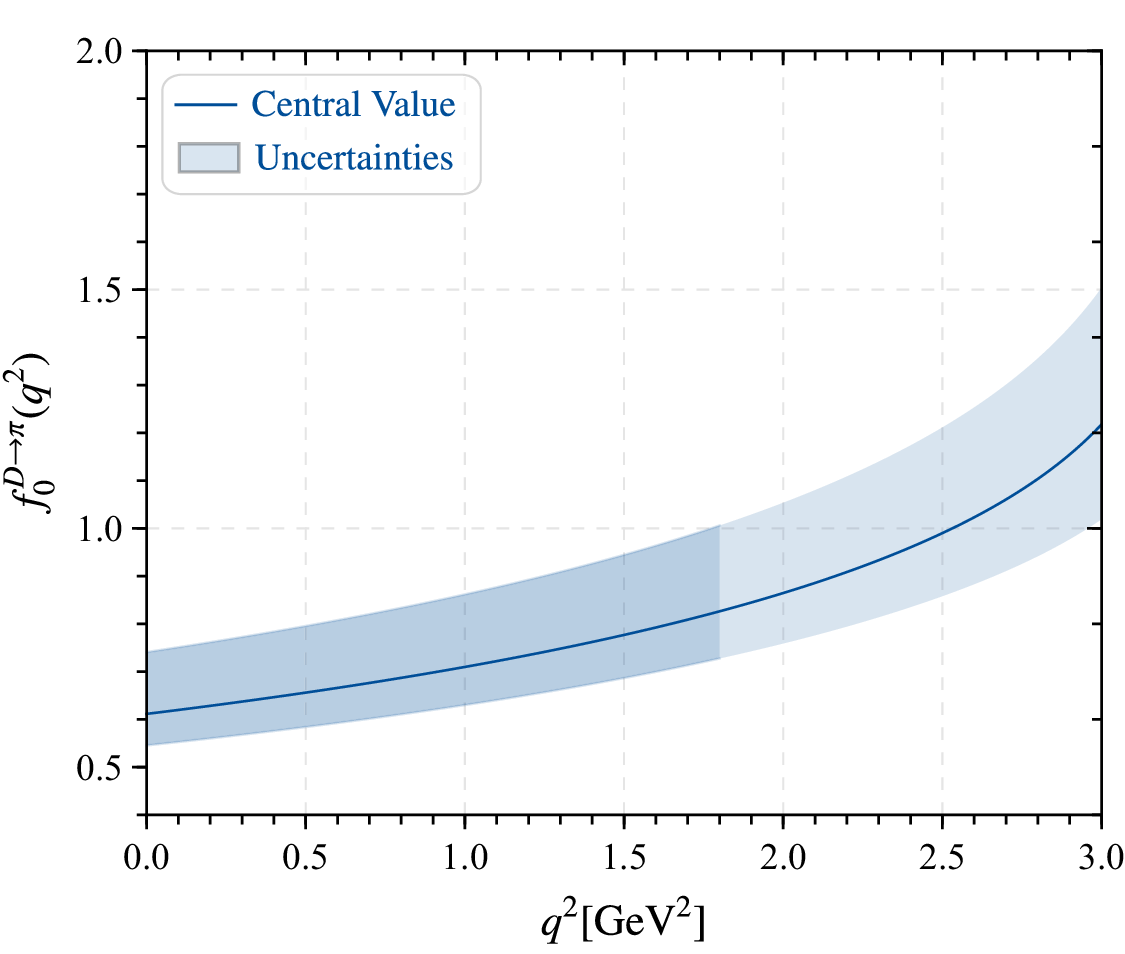}
\end{center}
\caption{$D\to \pi$ TFFs $f^{D\to \pi}_+(q^2)$ and $f^{D\to \pi}_0(q^2)$ in the whole physical region within uncertainties in this work. The results of other theoretical and experimental groups such as Belle~\cite{Belle:2006idb}, BESIII~\cite{BESIII:2015tql}, BaBar~\cite{BaBar:2014xzf},  CLEO~\cite{CLEO:2009svp} and LQCD~\cite{FermilabLattice:2004ncd}  collaborations are also shown as a comparison.}  \label{Fig:TFFfp}
\end{figure}

Based on the resultant pion twist-2 and 3 DA, we can then calculate the  $D\to \pi$ TFFs $f^{D\to \pi}_+ (q^2)$ and $f^{D\to \pi}_0 (q^2)$. The basic input parameters have been mentioned in the beginning of this section. So the main task is to determine the continuum threshold parameter $s_0$ and Borel windows $M^2$. Based on the basic ideas and processes of LCSR, we adopt the following three criteria.
\begin{itemize}
\item The continuum contributions are less than 30\% to the total results;
\item The contributions from higher-twist DAs are less than 5\%;
\item Within the Borel window, the changes of TFFs does not exceed 10\%
\item The continuum threshold $s_0$ should be closer to the squared mass of the first excited state $D$-meson.
\end{itemize}
So, we determined the continuum threshold parameter are $s_0^+ = \tilde s_0 = 5.2(3)~{\rm GeV^2}$. The Borel parameters are $M_+^2 = 8.0(5)~{\rm GeV^2}$ and $\tilde M^2=10.0(5)~{\rm GeV^2}$. After considering the errors coming from all the input parameters, we present the TFFs at large recoil region, {\it e. g. } $f^{D\to \pi}_+ (0) =  f^{D\to \pi}_0 (0)$ in Table~\ref{Tab:TFFfp0}. In which, the uncertainties are from the squared average of all the mentioned error sources.  As a comparison, we also present other theoretical and experimental predictions, such as the Belle~\cite{Belle:2006idb}, BESIII~\cite{BESIII:2015tql}, BES~\cite{BES:2004rav}, \textsc{Babar}~\cite{BaBar:2014xzf}, CLEO~\cite{CLEO:2009svp} Collaborations for the  experimental predictions, LQCD~\cite{FermilabLattice:2004ncd}, LCSR~\cite{Ball:2006yd} for the theoretical predictions respectively. Our results agree well with the BESIII, Belle, LQCD, \textsc{Babar}, LCSR and CLEO  predictions within errors.

\begin{figure}[t]
\begin{center}
\includegraphics[width=0.4\textwidth]{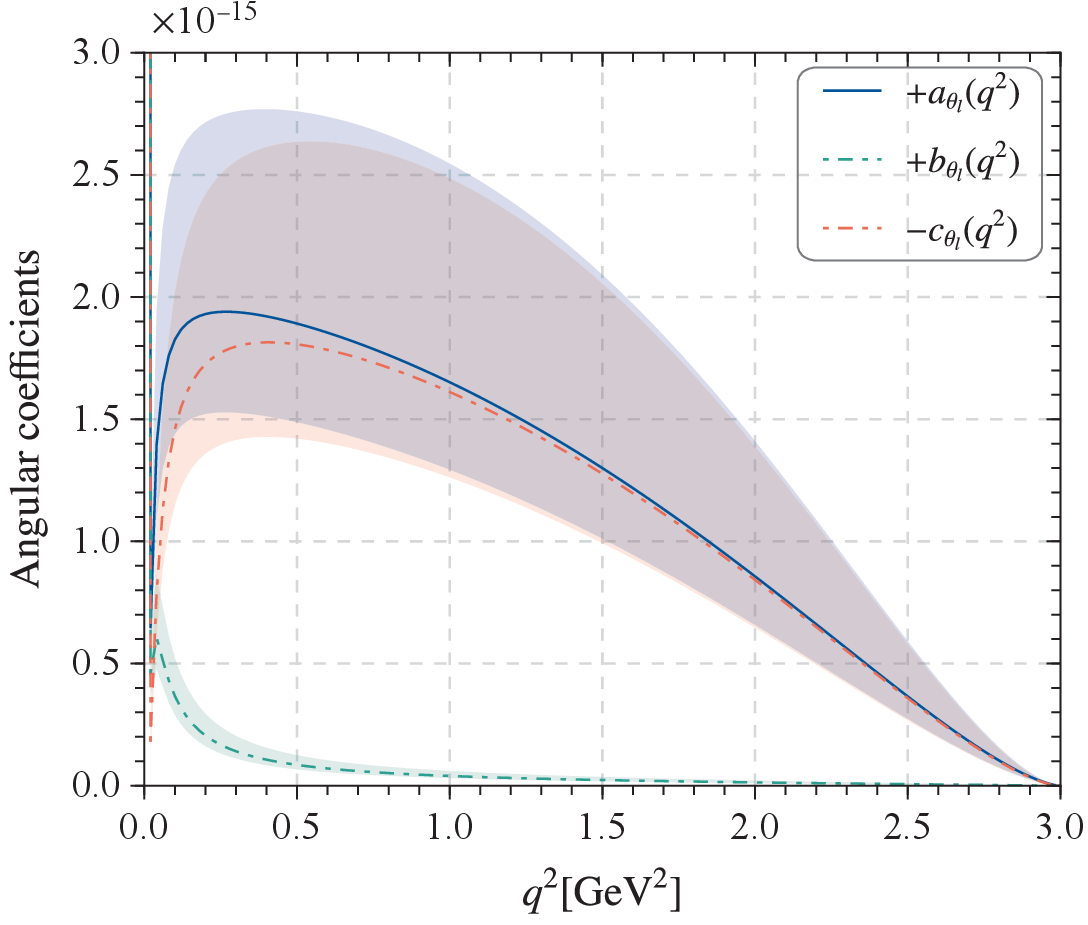}
\end{center}
\caption{The distribution of three $q^2$-dependent angular coefficient functions $a_{\theta_\ell}(q^2)$, $b_{\theta_\ell}(q^2)$ and $c_{\theta_\ell}(q^2)$ (in unit: $10^{-15}$), where the shaded bands stand for the uncertainties.}
\label{Fig:7}
\end{figure}

The physically allowable ranges for the $D\to \pi$ TFFs are $0 \leq q^2\leq q^2_{\rm max} = (m_D - m_\pi)^2\sim3~{\rm GeV^2}$. Theoretically, the LCSRs approach for $D \to \pi {\nu }{\bar\nu }$ TFFs are in low and intermediate $q^2$-regions, i.e. $0 \leq q^2\leq 1.8~{\rm GeV^2}$ of $\pi$-meson. One can extrapolate it to whole $q^2$-regions via a rapidly $z(q^2,t)$ converging the simplified series expansion (SSE), i.e. the TFFs are expand as~\cite{Bharucha:2015bzk}:
\begin{eqnarray}
f_+^{D\to \pi}(q^2) =\frac{1}{1-q^2/m_D^2}\sum_{k=0,1,2}{\beta _kz^k( q^2,t_0 )}
\end{eqnarray}
where $\beta_k$ are real coefficients and $z(q^2,t)$ is the function,
\begin{eqnarray}
z^k( q^2,t_0 ) =\frac{\sqrt{t_+-q^2}-\sqrt{t_+-t_0}}{\sqrt{t_+-q^2}+\sqrt{t_+-t_0}},
\end{eqnarray}
with $t_{\pm} = (m_{D} \pm m_{\pi})^2$ and $t_0=t_{\pm}(1-\sqrt{1-t_-/t_+})$. The SSE method possesses superior merit, which keeps the analytic structure correct in the complex plane and ensures the appropriate scaling, $f_+^{D\to \pi}(q^2)\sim 1/q^2$ at large $q^2$. And the quality of fit $\Delta$ is devoted to take stock of the resultant of extrapolation, which is defined as
\begin{eqnarray}
\Delta =\frac{\sum_t{| F_i(t) -F_{i}^{\mathrm{fit}}( t ) |}}{\sum_t{| F_i(t) |}}\times 100.
\end{eqnarray}
After making extrapolation for the TFFs $f_+^{D\to \pi}(q^2)$ to the whole physical $q^2$-region. Then, the behaviors of $D\to\pi$ TFFs in the whole physical region with respect to squared momentum transfer are given in Fig.~\ref{Fig:TFFfp}. In which, the darker band are the LCSR resluts of our prediction, while the lighter band are the SSE predictions. As a comparison, we also present the predictions from theoretical and experimental groups, such as the LQCD~\cite{FermilabLattice:2004ncd}, Belle~\cite{Belle:2006idb}, \textsc{Babar}~\cite{BaBar:2014xzf}, CLEO~\cite{CLEO:2009svp} and BESIII~\cite{BESIII:2015tql} collaborations are also presented. Our predictions have a good agreement with he experimental collaborations within errors. We found at the large squared momentum transfer, the predictions from LQCD have large gap with our results. Meanwhile, we also present the TFF $f_0^{D\to \pi}(q^2)$ in the lower panel of Fig.~\ref{Fig:TFFfp}. Furthermore, we present the behaviors of three angular coefficients functions $a_{\theta_\ell}(q^2)$, $b_{\theta_\ell}(q^2)$ and $c_{\theta_\ell}(q^2)$ uncertainties with $10^{-17}$-order level in Fig.~\ref{Fig:7}. The negative of $c_{\theta_\ell}$ is given for convenience to compare the three angular coefficients. As can be seen from the figure, the absolute values of $a_{\theta_\ell}(q^2)$ and $c_{\theta_\ell}(q^2)$ are very closer with uncertainties, and value for $b_{\theta_\ell}(q^2)$ is smaller than that of $a_{\theta_\ell}(q^2)$ and $-c_{\theta_\ell}(q^2)$.

\begin{figure}[t]
\begin{center}
\includegraphics[width=0.4\textwidth]{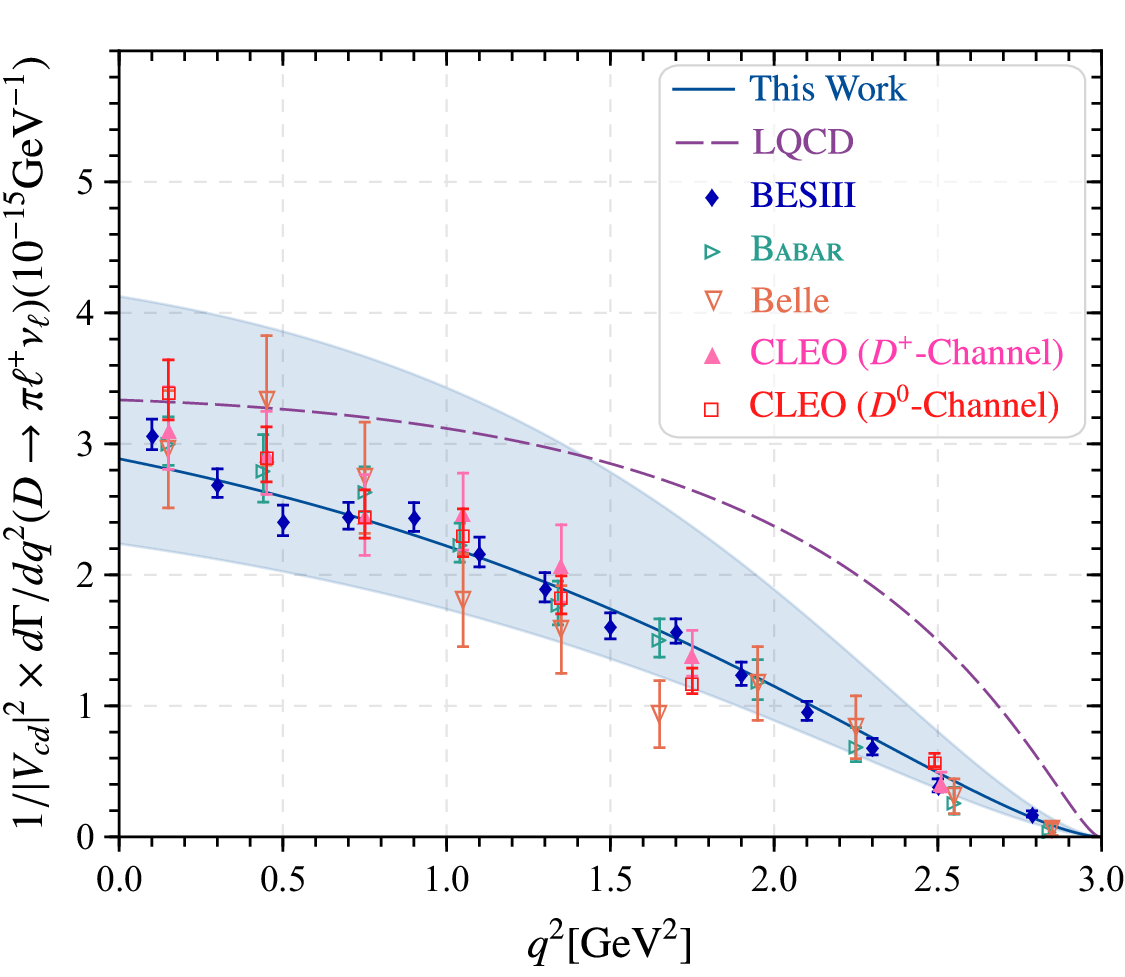}
\end{center}
\caption{The predictions of our calculation for $D\to \pi\ell^+\nu_\ell$ differential decay width within uncertainties (in unit: $10^{-15}$). Meanwhile, the results of other experimental groups such as  Belle~\cite{Belle:2006idb}, BESIII~\cite{BESIII:2015tql} \textsc{Babar}~\cite{BaBar:2014xzf}, CLEO~\cite{CLEO:2009svp} Collaborations and LQCD prediction~\cite{FermilabLattice:2004ncd} are also presented as a comparison.}\label{Fig:dGamma}
\end{figure}

Then our task is to calculate the decay width and branching fraction for the semileptonic decay processes $\bar D^0\to \pi^+ e \bar\nu_{e}$. By taking the $D\to \pi$ TFFs $f^{D\to \pi}_{+,0}(q^2)$ or angular coefficient function $a_{\theta_\ell}(q^2)$, $b_{\theta_\ell}(q^2)$, $c_{\theta_\ell}(q^2)$ into the expression of decay width, {\it e.g.} Eq.~\eqref{Eq:dGamma}, we present the curves of $D^+\to \pi e\nu_e$ differential decay width in Fig.~\ref{Fig:dGamma} (in unit: $10^{-15}$) with darker band are the uncertainties coming from all the input parameters. As a comparison, we also listed the predictions from  Belle~\cite{Belle:2006idb}, BESIII~\cite{BESIII:2015tql} \textsc{Babar}~\cite{BaBar:2014xzf}, CLEO~\cite{CLEO:2009svp} Collaborations and LQCD prediction~\cite{FermilabLattice:2004ncd} as a comparison. From the figure, we can see that the results of our predictions have agreement with the most experimental results within uncertainties, especially for the BESIII Collaboration.

Furthermore, after considering the $D^0$-meson lifetime $\tau(D^0) = 0.410(1)~{\rm ps}$ coming from PDG~\cite{PDGnew} and integrate the squared momentum transfer $q^2$, we can get the value of total branching fraction for the $D^0\to \pi^- e^+ \nu_e$, which are shown in Table~\ref{Tab:BF-Dpiev}. The error of our prediction are comes from all input parameters. To make a comparison, we also present the PDG~\cite{Chen:2007cn}, Belle~\cite{Belle:2006idb},
BESIII~\cite{BESIII:2015tql}, BES~\cite{BES:2004rav},
\textsc{Babar}~\cite{BaBar:2014xzf}, CLEO~\cite{CLEO:2009svp} and
LQCD~\cite{FermilabLattice:2004ncd} predictions. The value of our prediction have agreement with BESIII, PDG average value and LQCD results.

\begin{table}[b]
\footnotesize
\caption{The branching fraction (in unit $10^{-2}$) for semileptonic decay $\bar D^0\to\pi^+e\bar\nu_e$ of our predictions. Meanwhile, the theoretical and experimental results from other group are also given as a comparison.} \label{Tab:BF-Dpiev}
\begin{tabular}{lllll}
\hline
References~~~~ & Channel ~~~~~~~~~~~~& Predictions\\
\hline
This work & $\bar D^0\to\pi^+e\bar\nu_e$ & $0.308^{+0.155}_{-0.066}$\\
PDG~\cite{Chen:2007cn}&$\bar D^0\to\pi^+e\bar\nu_e$ & $0.281\pm0.190$\\
Belle~\cite{Belle:2006idb}&$D^0\to\pi^- \ell^+\nu_\ell$ & $0.255\pm0.019\pm0.016$\\
BESIII~\cite{BESIII:2015tql}&$D^0\to\pi^-e^+\nu_e$ & $0.295\pm0.004\pm0.003$\\
BES~\cite{BES:2004rav}&$D^0\to\pi^-e^+\nu_e$ & $0.330\pm0.130\pm0.030$\\
BaBar~\cite{BaBar:2014xzf}&$D^0\to\pi^-e^+\nu_e$ & $0.277\pm0.068\pm0.0092\pm0.037$\\
CLEO~\cite{CLEO:2009svp} &$D^0\to\pi^-e^+\nu_e$ & $0.288\pm0.008\pm0.003$\\
LQCD~\cite{FermilabLattice:2004ncd}&$D^0\to\pi^-\ell^+\nu_\ell$ & $0.316\pm0.025\pm0.062\pm0.033$\\
\hline
\end{tabular}
\end{table}
\begin{figure}[t]
\begin{center}
\includegraphics[width=0.4\textwidth]{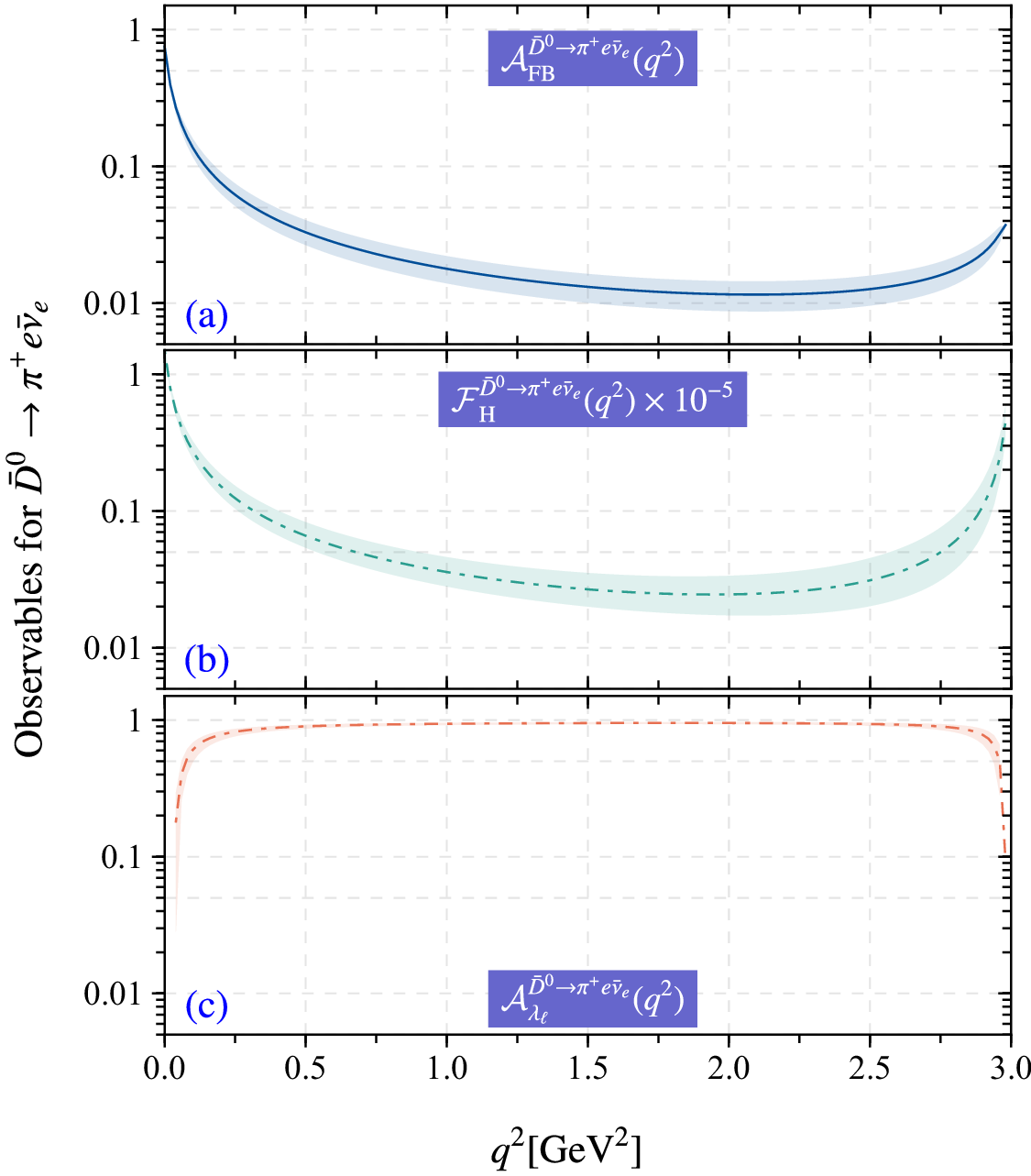}
\end{center}
\caption{Theory prediction for the three various items angular observables $\mathcal{A}_{\mathrm{FB}}^{
\bar D^0\to\pi^+ e\bar\nu_e}(q^2)$, $\mathcal{F}_{\mathrm{H}}^{\bar D^0\to\pi^+ e\bar\nu_e}(q^2)$, and $\mathcal{A}_{\lambda_\ell}^{\bar D^0\to\pi^+ e\bar\nu_e}(q^2)$. In which, the shaded bands stand for uncertainties.}  \label{Fig:Observables}
\end{figure}
On the other hand, the observables such as forward-backward asymmetries, $q^2$-differential flat terms and lepton polarization asymmetry, {\it e.g.} ${\cal A}_{\rm FB}^{\bar D^0\to\pi^+ e\bar\nu_e}(q^2)$, ${\cal F}_{\rm H}^{\bar D^0\to\pi^+ e\bar\nu_e}(q^2)$ and ${\cal A}_{\lambda_\ell}^{\bar D^0\to\pi^+ e\bar\nu_e}(q^2)$ can be obtained by using the resultant three angular coefficient, which can be found in our previous work~\cite{Tian:2023vbh}. The normalized forward-backward asymmetries and $q^2$ differential flat terms will vanish in the massless lepton limit to the SM, which is sensitive to beyond standard model (BSM). Meanwhile, the lepton polarization asymmetry is sensitive to helicity-violating new physics interactions. So, we display the three curves in Fig.~\ref{Fig:Observables} with (a), (b) and (c) panel respectively. The $q^2$ differential flat terms is in unit $10^{-5}$ order. The uncertainties of lepton polarization is very small that almost coincidence with central value.
\begin{figure}[t]
\begin{center}
\includegraphics[width=0.4\textwidth]{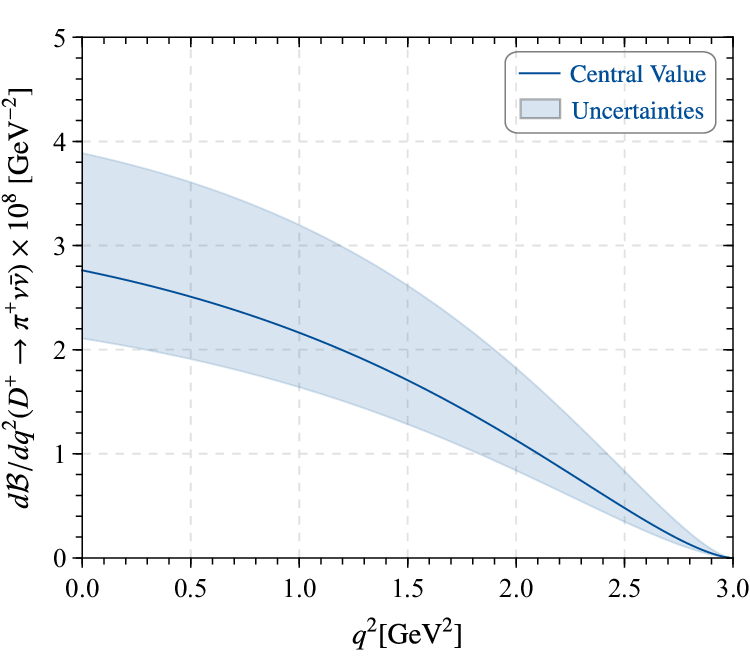}
\end{center}
\caption{The predictions of our calculation for $D^+\to \pi^+ \nu\bar\nu$ differential decay width within uncertainties (in unit: $10^{-8}$). Meanwhile, which is consistent with the upper limit of the experimental prediction results of BESIII~\cite{BESIII:2015tql} collaborations .}  \label{Fig:dBDpivv}
\end{figure}
For the next step, we can calculate the rare FCNC decays $D^+ \to \pi^+ \nu \bar{\nu}$. The input parameters are: Weinberg angle $\sin ^2{\theta _W}= 0.2312$, electromagnetic coupling constant $\alpha _{\rm em}$, running coupling constant $\alpha_s (1.1{\rm GeV})= 0.4256$, Fermi constant $G_F=1.166\times10^{-5}{\rm GeV^{-2}}$ and $C_F=4/3$. The curve for differential branching fraction of $D^+\to \pi^+\nu\bar\nu$ within uncertainties are exicibit in Fig.~\ref{Fig:dBDpivv}. Due to other theoretical predictions such as Ref.~\cite{Chen:2007cn} is in $10^{-7}$ order, we do not show here to make a comparison.

Then, we present the branching ratio of dineutrino decay mode $D\to \pi \nu\bar\nu$ in Table~\ref{Tab:BranchingDpivv} within uncertainties. The NSI~\cite{Mahmood:2014dkp}, the SM with Long Distance and short distance~\cite{Hewett:1995aw}, the LU, cLFC and General from Ref.~\cite{Golz:2021akk} are also given. As can be seen that our prediction is in the same order with NSIs prediction,  larger than the SM and lower than the LU, cLFC, General results. Meanwhile, our prediction also have the same order with BESIII result.

\section{Summary}\label{section:4}
In the paper, the rate decay process of $D^+\to \pi^+ \nu\bar\nu$ is studied within the framework of QCD sum rule approach. Firstly, we give expression of pion twist-2 and twist-3 DAs $\xi$-moments $\langle\xi_{2;\pi}^n\rangle|_\mu$ and $\langle\xi_{3;\pi}^{(p,\sigma),n}\rangle|_\mu$ by using the QCD sum rule approach under the background field theory. The first five terms to the twist-2 DA and first two terms of the $\xi$-moments are given in the processes scale $\mu_k = 1.1~{\rm GeV}$. In order to avoid the large uncertainties coming from Gegenbauer moment $a_n$, we constructed the LCHO model for twist-2, 3 LCDAs and determined the parameters by using the $\xi$-moments.

\begin{table}[t]
\footnotesize
\caption{The $D^+ \to \pi^+ {\nu }{\bar\nu}$ branching fraction in this work. Meanwhile, the theoretical and experimental results are given with comparison.}\label{Tab:BranchingDpivv}
\begin{tabular}{lrl}
\hline
References~~~~~~~~~~~~~~~~~~~~~~~~~~~~~~~~~~&& Result\\
\hline
This Work & &$1.85^{+0.93}_{-0.46}\times10^{-8}$ \\
NSIs~\cite{Mahmood:2014dkp}  & &$3.21\times10^{-8}$ \\
SM~\cite{Hewett:1995aw} (Long Distance)  & $<$ &$8\times10^{-16}$ \\
SM~\cite{Hewett:1995aw} (Short Distance) & &$3.9\times10^{-16}$ \\
LU~\cite{Golz:2021akk}                             & $<$ &$2.5\times10^{-6}$\\
cLFC~\cite{Golz:2021akk}                          & $<$ & $1.4\times10^{-5}$\\
General~\cite{Golz:2021akk}                      & $<$ & $ 5.2\times10^{-5}$\\
\hline
\end{tabular}
\end{table}

Secondly, we calculated the $D\to \pi$ vector and scalar TFFs within the QCD light-cone sum rule approach up to next-to-leading order accuracy. The value of TFF at large recoil region $f^{D\to \pi}_+ (0)$ is present in Table~\ref{Tab:TFFfp0}. After extrapolate the TFFs into whole $q^2$-region via simplified series expansion, we give the TFFs in Fig.~\ref{Fig:TFFfp}. The comparison with other predictions is made. Meanwhile, the three $q^2$-dependence angular coefficient functions $a_{\theta_\ell}(q^2)$, $b_{\theta_\ell}(q^2)$ and $c_{\theta_\ell}(q^2)$ are also given.

Furthermore, we analysis the differential decay width for $D^+\to \pi^+e\nu_e$ in Fig.~\ref{Fig:dGamma} and make a detailed comparison with BESIII, \textsc{Babar}, Belle, CLEO and LQCD predictions. The total branching fraction for $\bar D^0 \to \pi^+ e\bar\nu_e$ is given in Table~\ref{Tab:BF-Dpiev}. Then we give the forward-backward asymmetries, $q^2$ differential flat terms, lepton polarization asymmetry in Fig.~\ref{Fig:Observables}. Finally, the differential and total branching fraction of $D^+\to \pi^+\nu\bar\nu$ are given in Fig.~\ref{Fig:dBDpivv} and Table~\ref{Tab:BranchingDpivv}. Our prediction is in the region of the BESIII upper limits. With the stable operation of BESIII, the collision energy of BEPC-II collider will be greatly improved in the following seven years and more data results will be reported. We hope the $D^+\to \pi^+\nu\bar\nu$ channel prediction can be reported in the near further.\\

\acknowledgments
This work was supported in part by the National Natural Science Foundation of China under Grant No.12265010, No.12265009, the Project of Guizhou Provincial Department of Science and Technology under Grant No.ZK[2021]024 and No.ZK[2023]142.

\end{document}